\documentstyle[11pt,newpasp,twoside,epsf]{article}
\def\edcomment#1{\iffalse\marginpar{\raggedright\sl#1\/}\else\relax\fi}
\reversemarginpar

\newcommand{\apg}{\:^{>}_{\sim}\:}
\newcommand{\apl}{\:^{<}_{\sim}\:}
\newcommand{\cmjj}{\mbox{${\rm cm^{-2}}$}}
\newcommand{\etal}{et al.}
\newcommand{\hI}{\mbox{${\rm H\ I}$}}
\newcommand{\kms}{\mbox{km\ s${^{-1}}$}}
\newcommand{\lya}{\mbox{${\rm Ly}\alpha$}}
\newcommand{\lyb}{\mbox{${\rm Ly}\beta$}}
\newcommand{\civ}{\mbox{${\rm C\,IV}$}}
\newcommand{\cii}{\mbox{${\rm C\,II}$}}
\newcommand{\oi}{\mbox{${\rm O\,I}$}}
\newcommand{\ovi}{\mbox{${\rm O\,VI}$}}
\newcommand{\nii}{\mbox{${\rm N\,II}$}}
\newcommand{\nv}{\mbox{${\rm N\,V}$}}
\newcommand{\siiv}{\mbox{${\rm Si\,IV}$}}
\newcommand{\siii}{\mbox{${\rm Si\,II}$}}
\newcommand{\siiii}{\mbox{${\rm Si\,III}$}}

\newcommand{\ibid}{\underline{\makebox[0.5in]{}}.}

\begin{document}
\title{Properties of a Chemically Enriched \lya\ Absorption System at 
$z=0.167$ and Implications for the Total Baryons in Extended Gaseous Envelopes 
Around Galaxies}
\author{Hsiao-Wen Chen and Jason X. Prochaska}
\affil{Carnegie Observatories, 813 Santa Barbara St., Pasadena, CA 91101}
\author{Kenneth M. Lanzetta}
\affil{Physics and Astronomy Department, SUNY at Stony Brook, Stony Brook,
NY 11794-3800}


\begin{abstract}

  We present a detailed analysis of the chemical abundances, ionization state, 
and origin of a partial Lyman limit system ($N(\hI) \approx 10^{16}$ \cmjj) at 
low redshift ($z=0.167$ towards PKS0405$-$1219).  We also present the first 
estimate of the total baryons in extended gaseous envelopes around galaxies
based on a simple photoionization model for the \lya\ absorbers.  We find that
extended gaseous envelopes can explain between 5--20\% of the total baryons 
predicted by the big bang nucleosynthesis model.

\end{abstract}

\section{Background}

  Various studies have shown that the ``forest'' of \lya\ absorption lines 
observed in the spectra of background QSOs are a sensitive probe of the 
physical conditions of intervening gas (e.g.\ Rauch 1998).  Comparisons of 
galaxies and QSO absorption systems along common lines of sight, therefore,
allow us not only to determine the origin of QSO absorption systems but also to
study properties of tenuous gas around galaxies at large distances.  Over the 
past several years, studies of faint galaxies in QSO fields have demonstrated 
that galaxies of a wide range of luminosity and morphological type possess 
extended gaseous envelopes of radius $\sim 180\ h^{-1}$ kpc and that known 
galaxies of known gas cross section can explain most and perhaps all of \lya\ 
absorption systems of $N(\hI) \apg 10^{14}$ \cmjj (Lanzetta \etal\ 1995; Chen 
\etal\ 1998, 2001).  A more detailed review of the relationship between \lya\ 
absorption systems and galaxies is presented by Lanzetta in these proceedings. 
Here we summarize the main reasons that support this conclusion as following:

  1. The galaxy--absorber cross-correlation function exhibits a strong 
amplitude on velocity scales of $\apl$ 250 \kms\ and impact parameter scales
of $\apl 200\ h^{-1}$ kpc, and becomes diminishingly small at impact parameters
beyond $200 \ h^{-1}$ kpc on all velocity scales (Lanzetta, Webb, \& Barcons 
1997).  This indicates that \lya\ absorption systems strongly cluster around
galaxies on dynamical scales of individual galaxies, rather than on scales of
large filamentary structures.

  2. There exists a strong anti-correlation between \lya\ absorption equivalent
width and galaxy impact parameter.  Namely, galaxies at larger impact
parameters are found to be associated with weaker \lya\ absorption systems.  
Including the intrinsic luminosities of individual galaxies as an additional 
scaling factor further improves the anti-correlation between \lya\ absorption 
equivalent width and galaxy impact parameter, indicating that the extent of 
tenuous gas around galaxies scales with individual galaxy luminosity/mass (Chen
\etal\ 1998, 2001).  

  3. Almost all galaxies that are at $\delta v \apg 3000$ \kms away from the 
background QSOs and at $\rho \apl 200\ h^{-1}$ kpc produce associated \lya\ 
absorption lines, indicating that the covering factor of tenuous gas around 
galaxies is approximately unity (Lanzetta \etal\ 1995; Chen \etal\ 1998, 2001).
On the contrary, almost all galaxies that are in the vicinities of the 
background QSOs ($\delta v \apl 3000$ \kms and $\rho \apl 200\ h^{-1}$ kpc) do 
not produce associated \lya\ absorption lines to within sensitive upper limits,
consistent with the picture that tenuous gas around these galaxies are highly 
ionized due to an enhanced ionizing radiation intensity from the background 
QSOs (Pascarelle \etal\ 2001).

  By combining results of previous studies of extended Mg\,II, C\,IV, and 
\lya\ gas around galaxies (derived based on analyses of QSO absorption 
systems), we can establish a schematic picture of the structure of extended gas
around galaxies: A typical $L_*$ galaxy is surrounded by high column density 
gas of radius a few tens kpc (e.g.\ van Gorkom 1993), which, because of its 
high density, remains mostly neutral.  The gas becomes more ionized as the gas
density decreases with increasing galactic radius.  Low-ionization species,
such as singly ionized magnesium Mg\,II, become the dominant observational 
signature out to $\approx$ 50 $h^{-1}$ kpc (e.g.\ Bergeron \& Boiss\'e 1991).  
As the gas density continues to decline at larger galactic radii, the gas 
becomes still more highly ionized.  High-ionization species, such as triply 
ionized carbon \civ, become the dominant observational signature out to 
$\approx$ 100 $h^{-1}$ kpc, at neutral hydrogen column densities $N(\hI) 
\approx 10^{16}$ \cmjj\ (Chen, Lanzetta, \& Webb 2001).  The tenuous gas 
continues to extend to at least $\approx$ 180 $h^{-1}$ kpc, at neutral hydrogen
column densities at least as low as $N(\hI) \approx 3\times 10^{14}$ \cmjj\ 
(Chen \etal\ 1998, 2001).  This picture applies to galaxies of a wide range of 
luminosity and morphological type.

  There is, however, a lack of understanding of the physical properties of 
extended gas at large galactic distances, such as number density, ionization 
state, and metallicity, because of limited information regarding kinematics and
metallicity of the gas.  The primary difficulty arises in acquiring a 
high-quality QSO spectrum of wide UV spectral coverage.  Here we present an
analysis of the chemical abundances, ionization state, and origin of a partial 
Lyman limit system (LLS; $N(\hI) \approx 10^{16}$ \cmjj) at low redshift ($z =
0.167$ towards PKS0405$-$1219).  Combining the results of our analysis for the 
$z = 0.167$ system with those for high-redshift \lya\ absorption systems 
obtained in the literature, we also present the first estimate of the total 
baryons in extended gaseous envelopes around galaxies based on a simple 
photoionization model.

\section{The Partial Lyman Limit System at $z=0.167$}

  The absorption system at $z = 0.167$ toward PKS0405$-$1219 ($z_{em}=0.5726$)
was first identified in an HST/FOS spectrum with a rest-frame \lya\ absorption 
equivalent width of 0.65 \AA\ and a rest-frame \civ\ absorption equivalent
width of 0.44 \AA.  Subsequent echelle spectroscopy with HST/STIS further 
revealed absorption features produced by H$^0$ (\lyb), C$^+$, N$^+$, O$^0$, 
Si$^+$, Si$^{++}$, Si$^{+3}$, possibly Fe$^+$ and Fe$^{++}$; and most 
interestingly absorption doublets produced by N$^{+4}$ and O$^{+5}$ at the 
redshift of the absorption system.  We estimated the physical properties of
the absorption system based on both photoionization and collisional ionization 
models, using the relative abundance ratios between different transitions.
Detailed analysis has been presented in Chen \& Prochaska (2000).  We summarize
the results here.

  First, all the ions except N$^{+4}$ and O$^{+5}$ showed consistent profile 
signatures that trace the partial LLS, while the \nv\ and \ovi\ doublets 
appeared to be broad with velocity centroids blue-shifted by $\approx$ 30 
\kms\ from the other ions.  The differences in kinematic signatures strongly 
indicate that the \nv\ and \ovi\ doublets and the partial LLS traced by the 
other ions do not arise in the same regions (see Figure 1 in Chen \& Prochaska 
2000).

  Second, metals that are associated with the partial LLS showed a wide range 
of ionization state.  We present the column density measurements in Table 1, 
which lists the ions, the corresponding rest-frame absorption wavelength 
$\lambda_0$, and the estimated ionic column densities $\log\,N$ together with 
the associated errors in the first three columns.  The comparable abundances 
between various Si ions indicate that the partial LLS is photoionized.  

\begin{center}
\begin{tabular}{p{0.5in}lrrr}
\multicolumn{5}{c}{Table 1. Measurements of Column Densities, }\\ 
\multicolumn{5}{c}{Doppler Parameters, and Abundances} \\
\hline
\multicolumn{1}{c}{Species} & \multicolumn{1}{c}{$\lambda_0$} & 
\multicolumn{1}{c}{log $N$ (cm$^{-2}$)$^a$} & \multicolumn{1}{c}{$b$ (\kms)}
& \multicolumn{1}{c}{[X/H]$^b$} \\
\hline
\hline
 \hI\    \dotfill & 1025.72 & $>$ 15.7 & $34.7 \pm 2.5$ & ... \\
                  & 1215.67 & $>$ 15.7 & ...  & ... \\
 \cii\   \dotfill & 1036.79 & $>14.10$ & ...  & ... \\
                  & 1334.53 & 14.27 $\pm$ 0.09 & ...  & $-0.33$ \\
 \nii\   \dotfill & 1083.99 & $>14.25$ & $11.6 \pm 1.0$ &  $0.27$ \\
 \nv\    \dotfill & 1238.82 & 13.84 $\pm$ 0.07 & $50.2 \pm 7.0$ & ... \\
                  & 1242.80 & 13.91 $\pm$ 0.06 & ...  & ... \\
 \oi\    \dotfill & 1302.17 & 13.68 $\pm$ 0.14 & ...  & $0.25$ \\
 \ovi\   \dotfill & 1031.93 & 14.67 $\pm$ 0.16 & $72.6 \pm 4.9$ & ... \\
                  & 1037.62 & 14.76 $\pm$ 0.07 & ...  & ... \\
 \siii\  \dotfill & 1190.42 & 13.22 $\pm$ 0.07 & $9.9 \pm 0.9$  & ... \\
                  & 1193.29 & 13.29 $\pm$ 0.05 & ...  & ... \\
                  & 1260.42 & $>13.16$ & ...  & ... \\
                  & 1304.37 & 13.40 $\pm$ 0.12 & ...  & $-0.37$ \\
 \siiii\ \dotfill & 1206.50 & $>13.33$ & ...  & $>-0.76$  \\
 \siiv\  \dotfill & 1393.76 & 13.18 $\pm$ 0.05 & ...  & ... \\
                  & 1402.77 & 13.49 $\pm$ 0.07 & ...  & $-0.45$  \\ 
\hline
\multicolumn{5}{l}{$^a$Lower limits indicate that the lines are saturated.} \\
\multicolumn{5}{l}{$^b$$[{\rm X}/\,{\rm H}]$ is defined as 
$\log\,[N({\rm X})/\,N({\rm H})] - \log\,[{\rm X}/\,{\rm H}]_\odot$.} \\
\end{tabular}
\end{center}

  Third, comparisons of the relative abundances of the Si ions and predictions 
based on a series of photoionization models calculated using the CLOUDY 
software package (Ferland 1995) showed that the ionization parameter, which is 
the ratio of incident ionizing photons at 912 \AA\ to the total hydrogen number
density, $U\equiv\phi_{912}/\,cn_{\rm H}$, was constrained to be $\log\,U = 
-2.64 \pm 0.07$.  In addition, the observed abundances of the N$^{+4}$ and 
O$^{+5}$ ions cannot be explained by the same photoionization models (Figure 
1).  For an extreme upper limit $\log\,U=-2.2$, both \ovi\ and \nv\ were 
predicted to be at least 0.4 dex less abundant than \siiv, while our 
measurements showed that \ovi\ and \nv\ are at least 0.4 dex more abundant than
all the Si ions.  This result, together with their kinematic signature, 
indicates that these ions arise in a different region and suggests that they 
may be collisionally ionized.  

\begin{figure}[t]
\includegraphics{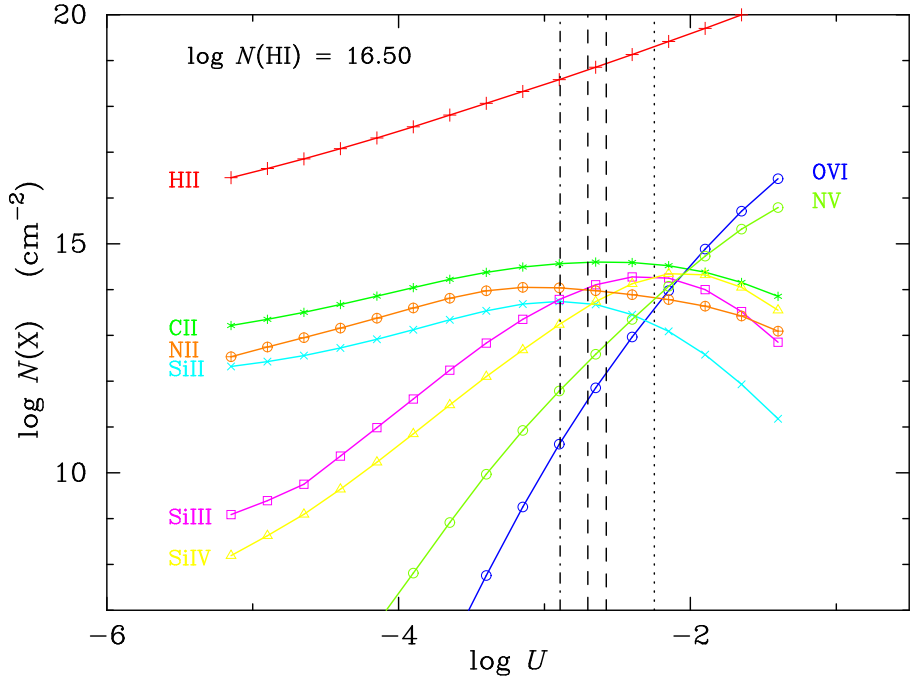}
\vspace{3.4in}
\caption{Predicted relative abundances of various ions X for a plane-parallel 
slab of absorbing gas of solar metallicity based on a photoionization analysis
using CLOUDY.  The dashed lines indicate the plausible range of the ionization 
parameter $U$ determined using the observed abundance ratios between Si${^+}$ 
and Si${^{+3}}$ for the system at $z=0.167$; the dotted and dot-dashed lines
indicated the lower and upper limits of $U$ determined based on the ratios of
Si$^{++}/\,$Si$^+$ and Si$^{+}/\,$Si$^{++}$.}
\end{figure}
 
  Fourth, the temperature of the partial LLS was found to be $T\approx 7.4 
\times 10^4$ K with a 1 $\sigma$ lower limit being $T\approx 2.7 \times 10^4$ K
and 1 $\sigma$ upper limit being $T\approx 1.5 \times 10^5$ K, using the 
Doppler parameters $b$ estimated from a Voigt profile analysis (column 4 of 
Table 1) for the unsaturated lines produced by Si$^+$ and N$^+$.  Because the 
cooling function is the most effective at $T\approx 10^5$ K (see e.g.\ Spitzer
1978), the temperature of the partial LLS is more likely to be a few times 
$10^4$ K. 

  Fifth, assuming that the more highly ionized component traced by the O$^{+5}$
and N$^{+4}$ ions is collisionally ionized, we estimated the temperature to be
$T = (2.6\pm 0.1) \times 10^5$ K by comparing the observed \ovi\ and \nv\ 
column density ratio with predictions based on a series of collisional 
ionization models presented in Shapiro \& Moore (1976).  In order for the hot 
gas to remain at this temperature, we further estimated the mass of the galaxy 
or galaxy group that supports the extended hot gas to be $\approx 1.5 \times 
10^{12} M_\odot$, assuming a simple isothermal profile of a half-mass radius 
$r_h \approx 0.35$ Mpc.  By requiring a cooling time longer than the dynamical
time, we estimated the number density of the absorbing gas should be $\apl 3
\times 10^{-5}$ cm$^{-3}$.

  Finally, we estimated the elemental abundances of the partial LLS using the 
ionization fraction correction calculated from the CLOUDY model for an 
ionization parameter $\log\,U=-2.64$.  In addition, we also estimated the 
chemical abundance using the $[\oi/\,\hI]$ column density ratio, which is a 
direct estimate of the oxygen elemental abundance in regions where the resonant
charge-exchange reaction between oxygen and hydrogen becomes dominant.  The 
results are shown in the last column of Table 1 for an \hI\ column density 
$\log\,N(\hI)=16.5$.  It shows that the partial LLS would have $1/10$ solar 
abundance determined for carbon and silicon, $1/5$ solar abundance for 
nitrogen, and at least $1/2$ solar for oxygen.  Increasing (decreasing) 
$N(\hI)$ by 0.5 dex decreases (increases) the abundance measurements by 0.5 
dex.


  To summarize, the \lya\ absorption system at $z=0.167$ is a chemically 
enriched partial LLS that exhibits a wide range of ionization state and is 
likely to be embedded in a more wide-spread, collisionally ionized hot gas that
gives rise to the absorption features of O$^{+5}$ and N$^{+4}$.  A summary of 
the physical parameters of the warm and hot gas that are associated with this 
system is presented in Table 2 of Chen \& Prochaska (2000).  Ground-based 
galaxy surveys (Spinrad \etal\ 1993; Ellingson \& Yee 1994) of the field 
surrounding PKS0405$-$1219 have found an elliptical galaxy at $z=0.1667$ and 
impact parameter $\rho = 74.9\ h^{-1}$ kpc with a rest-frame $K$-band 
luminosity $L_K = 1.20\ L_{K_*}$ and (2) a spiral galaxy at $z=0.1670$ and at 
$\rho = 62.8 \ h^{-1}$ kpc with $L_K = 0.02 \ L_{K_*}$.  By carefully examining
deep optical and near-infrared images, we find that if there are unidentified 
dwarf galaxies at the absorber redshift and closer to the QSO line of sight, 
then they cannot be brighter than $0.02\ L_{K_*}$.  By applying the luminosity 
scaled anti-correlation between \lya\ absorption strength and galaxy impact 
parameter presented in Chen \etal\ (2001), we find that the elliptical galaxy 
alone can easily explain the amount of absorption.  While it is not uncommon to
find tenuous gas at large galactic distance ($\rho \sim 100 \ h^{-1}$ kpc) 
around nearby elliptical galaxies (e.g.\ Fabbiano, Kim, \& Trinchieri 1992) or 
in groups of galaxies (e.g.\ Mulchaey \etal\ 1996), it is surprising to find 
metal-enriched, high-column density gas at this large distance.  The spectral 
characteristics of the elliptical galaxy exhibit signs of recent star formation
(Spinrad \etal\ 1993), therefore the physical process that initiated the recent
star formation might be responsible for transporting metals to large galactic 
distance.

\section{Total Baryons in Extended Gaseous Envelopes Around Galaxies}

  Big bang nucleosynthesis (BBN) makes specific predictions for the light 
element abundances (D, $^3$He, $^4$He, and $^7$Li) as a function of the nucleon
to proton ratio $\eta$ (e.g.\ Schramm \& Turner 1998).  Abundance measurements 
of these elements determine $\eta$ and therefore reveal the mean baryon density
$\Omega_b$ in the universe when combined with measurements of the cosmic 
background radiation.  Measurements of the deuterium abundance along three
sightlines toward high-redshift QSOs have yielded a precise estimate of the 
baryon density of $\Omega_b\,h^2=0.02$ (e.g.\ Burles, Nollett, \& Turner 2001).
Combined data from various cosmic microwave background experiments have also
shown exciting agreement in the estimate of $\Omega_b$ (e.g.\ Wang, Tegmark, \&
Zaldarriaga 2001).  The consistent results in $\Omega_b$ greatly increase the
confidence in our current understanding of the formation of the universe.  
While all the baryons may be accountable by the \lya\ forest at high redshifts 
(Rauch \etal\ 1997), it has, however, been shown that the sum of all the 
baryons seen in different environment only makes up $1/3$ of the total baryons 
predicted by BBN (Fukugita, Hogan, \& Peebles 1998).  Finding the missing 
baryons is therefore crucial for establishing a complete picture of the 
formation and evolution of the universe.

  Fukugita \etal\ (1998) attributed the ``missing baryons'' to the hot gas 
around galaxy groups with temperatures $\approx 10^5$ K, a temperature range 
that is out of reach by existing X-ray facilities but may be probed using the 
\ovi\ absorption systems observed in the spectra of background QSOs (Mulchaey 
\etal\ 1996).  Results of hydrodynamic simulations also suggested that 
30\%--40\% of the total baryons are in ionized gas of temperatures between 
$10^5$ and $10^7$ K at low redshifts (Cen \& Ostriker 1999; Dav\'e \etal\ 
2001).  Recent analysis of intervening \ovi\ absorption systems indeed showed 
evidence that supports a significant baryon reservoir in the hot gas probed by
the \ovi\ absorbers (Tripp, Savage, \& Jenkins 2000).  There is, however, a
large uncertainty due to poor statistics in the line densities of the \ovi\ 
absorption systems and unknown gas properties such as metallicity and 
ionization fraction of the gas.  In addition, some fraction of the \lya\ forest
have associated \ovi\ absorption lines, a simple addition of the total baryon 
contributions from the two species will result in a double-counting error (see 
the contribution by Tripp in these proceedings).

  Because extended gaseous envelopes are a common and generic feature of 
galaxies of a wide range of luminosity and morphological type (Chen \etal\ 
1998, 2000), some fraction of the total baryons must reside in the gaseous
envelopes of individual galaxies that have not been counted in previous 
measurements.  Here we present the first estimate of the total baryons in 
tenuous gas around galaxies on the basis of known galaxies that are represented
by a galaxy luminosity function and of known gaseous extent of these galaxies.
Our analysis takes advantage of better known galaxy statistics and is based on
hydrogen directly.  It requires an accurate estimate of the ionization fraction
of the tenuous gas, but it does not require knowledge of the metallicity of the
gas.

  To determine the ionization fraction of an absorber requires high-resolution,
high-sensitivity QSO spectroscopy to identify associated metal lines in the 
rest-frame UV spectral range.  A handful of \lya\ absorption systems have so 
far been studied extensively for the ionization state of the absorbing gas 
(Bergeron \etal\ 1994; Lopez \etal\ 1999; Prochaska 1999; Prochaska \& Burles 
1999; Rauch \etal\ 1999), in addition to the partial LLS at $z=0.167$.  All of
these absorption systems are best explained by a simple photoionization model,
in which the ionization fraction of the gas clouds only depends on the 
ionization parameter $U$.  Adopting a best estimate of the ionizing flux 
intensity $J_{912}$ for a gas cloud of known geometry, we can therefore infer
the ionization fraction for a \lya\ absorber of a given neutral hydrogen column
density.

  Figure 2 shows the predicted ionization fraction $X\equiv N({\rm H}^+)/
N({\rm H})$ as a function of neutral hydrogen column density $N(\hI)$ (curves) obtained using CLOUDY for a range of 
$J_{912}$.  Estimates of a handful \lya\ absorption systems obtained from the
literature are shown in closed points with the two $z<1$ systems highlighted in
circles (the $z=0.167$ system described in \S\ 2 and the $z=0.79$ system
studied by Bergeron \etal\ 1994).  We have also included a clumping factor $C =
\sqrt{\langle n_{\small{\rm H\,I}}^2\rangle/\langle n_{\small{\rm H\,I}}
\rangle^2}$, where $n_{\small{\rm H\,I}}$ is neutral hydrogen number density, 
in the models to account for the clumpiness of the gas.  Comparisons of various
models show that a higher $J_{912}$ would imply a lower $C$ and vice versa for
a system with known $X$ and $N(\hI)$.  Given that the background ionizing 
radiation intensity evolves rapidly with redshift (e.g.\ Haardt \& Madau 1996),
we determine a best-fit range in $J_{912}$ and $C$ based only on the two $z<1$
points.  We find that it is {\em unlikely} to have $J_{912}\apg 10^{-23}$ ergs
sec$^{-1}$ cm$^{-2}$ Hz$^{-1}$ Sr$^{-1}$ and $C\apg 10$ for these low-redshift
absorbers.  If we adopt the best estimated $J_{912}$ obtained from the QSO 
proximity effect measurements (Pascarelle \etal\ 2001; Scott \& Bechtold in 
these proceedings), 0.5--1 $\times 10^{-23}$ ergs sec$^{-1}$ cm$^{-2}$ 
Hz$^{-1}$ Sr$^{-1}$, then we find a clumping factor of order unity.

\begin{figure}[t]
\includegraphics{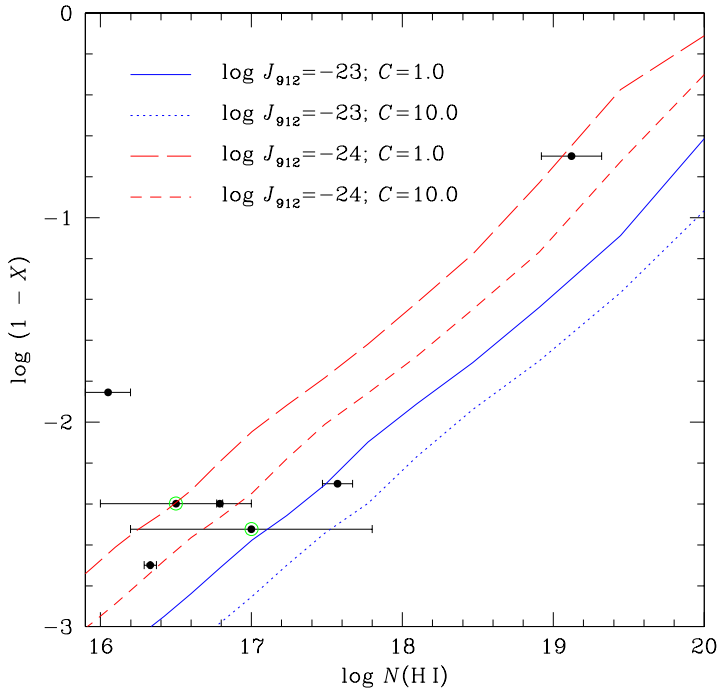}
\vspace{3.5in}
\caption{Comparisons of estimated ionization fraction for seven \lya\ 
absorption systems and predictions based on a series of photoionization models.
Points with circles indicate the two absorbers at $z<1$.  Curves indicate
predicted ionization fraction as a function of neutral hydrogen column density
$N(\hI)$ for a range of ionizing flux intensity $J_{912}$ at 912 \AA\ and a
range of neutral hydrogen clumping factor $C$.}
\end{figure}

\begin{figure}[t]
\includegraphics{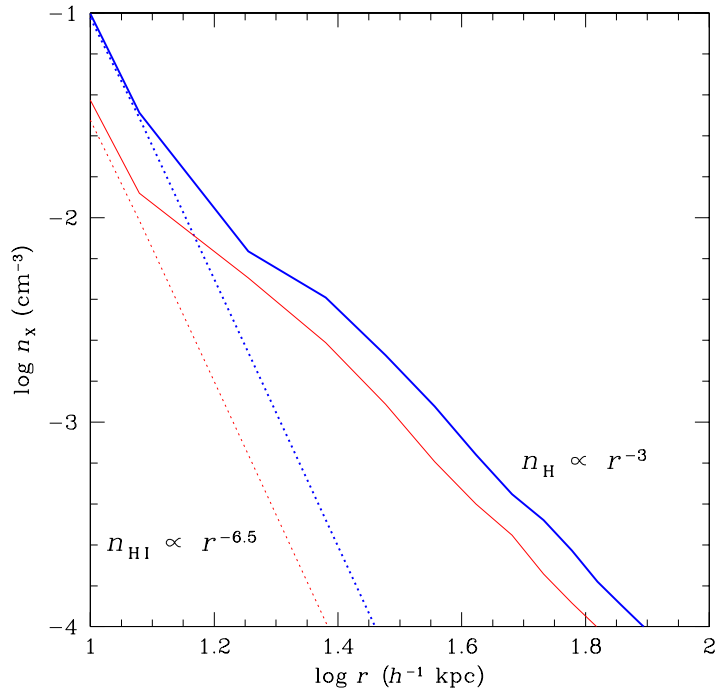}
\vspace{3.5in}
\caption{Hydrogen density profiles of a typical $L_*$ galaxy, assuming $J_{912}
= 10^{-23}$ ergs sec$^{-1}$ cm$^{-2}$ Hz$^{-1}$ Sr$^{-1}$.  Thick curves are 
for $C=1$; thin curves are for $C=10$.  Solid curves indicate the estimated 
total hydrogen profile; dotted curves indicate the neutral hydrogen profiles.
It shows that extended gas is mostly neutral at $r \apl 15\ h^{-1}$ kpc and 
therefore neutral hydrogen traces the total hydrogen gas profile.  At $r \apg 
15\ h^{-1}$ kpc, extended gas becomes highly ionized.  While the neutral 
hydrogen gas profile appears to be as steep as $n_{\small{\rm H\,I}} \propto 
r^{-6.5}$, the total gas density profile is well represented by $n_{\small{\rm 
H}} \propto r^{-3}$.}
\end{figure}

  To estimate the total baryons in extended gaseous envelopes of galaxies, we 
first determine the total hydrogen number density profiles of individual 
galaxies.  Adopting the neutral hydrogen column density profile in Chen \etal\ 
(2001),
\begin{equation}
N_{\rm H\,I}(\rho,L_B)=3.63\times 10^{21}\left(\frac{\rho}{10\,{\rm kpc}}\right)^{-5.56\pm 0.42}\left(\frac{L_B}{L_{B_*}}\right)^{2.56\pm 0.43} {\rm cm}^{-2},
\end{equation}
we derive the underlying neutral hydrogen number density profile, according to
\begin{equation}
n_{\small{\rm H\,I}}(r,L_B)=-\frac{1}{\pi C} \int^\infty_r\frac{dN_{\small{\rm H\,I}}(\rho,L_B)}{d\rho}\frac{d\rho}{\sqrt{\rho^2-r^2}}.
\end{equation}
To derive a total hydrogen number density profile, we apply proper ionization 
corrections determined from the CLOUDY photoionization calculations for gas of
a given $n_{\small{\rm H\,I}}$.  The results are shown in Figure 3 for extended
gas around a typical $L_*$ galaxy.  It shows that extended gaseous profiles may
be well described by a power-law $r^{\alpha}$ with $\alpha\approx -3$, which is
steeper than the one expected from an isothermal equilibrium model but agrees 
well with expectation from a dissipational formation scenario (Silk \& Norman 
1981).

\begin{figure}[t]
\includegraphics{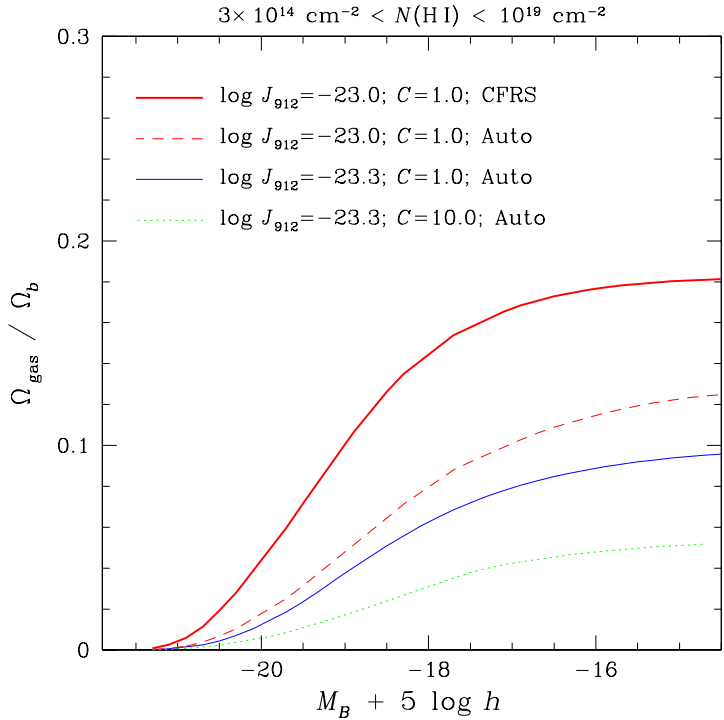}
\vspace{3.55in}
\caption{Cumulative contribution of baryons in extended gaseous envelopes of 
galaxies of different luminosity.  The calculations are for regions that 
correspond to $3\times 10^{14}\ {\rm cm}^{-2}\apl N(\hI)\apl 10^{19}\ {\rm 
cm}^{-2}$.}
\end{figure}

  Next, we calculate the cumulative contribution of baryons in extended gaseous
envelopes from galaxies of different luminosity according to

\begin{equation}
\Omega_b\equiv\frac{\rho_b}{\rho_c}=\frac{4\pi\mu m_{\rm H}}{\rho_c}\int\phi\left(\frac{L_B}{L_{B_*}}\right)\,d\left(\frac{L_B}{L_{B_*}}\right)\int_{r1}^{r2}r^2\,n_{\rm H}(r,L_B)\,dr,
\end{equation}
where $\mu$ is the mean atomic weight, $m_{\rm H}$ is the proton mass, $\rho_c$
is the present-day critical mass density, and $\phi$ is the galaxy luminosity
function.  The limits of the integral $(r_1, r_2)$ correspond to the column
density interval of interest here with the lower limit $N(\hI)=3\times 10^{14}
\ {\rm cm}^{-2}$ set by the sensitivity of the galaxy survey from which 
equation (1) was determined.

  Given the uncertainties in the galaxy luminosity function measurements, we
calculate equation (3) using results from the CFRS (Lilly \etal\ 1995) and 
Autofib (Ellis \etal\ 1996) surveys.  The results are shown in Figure 4.  It 
shows that extended gaseous envelopes of luminous galaxies ($M_{\rm B}\apl 
M_{{\rm B}_*}$) contribute most of the baryons, as indicated by the flattening 
of the curves toward fainter magnitudes beyond $L_*$.  The luminosity function 
of Lilly \etal\ has a higher normalization in galaxy number density and a 
shallower faint-end slope than that of Ellis \etal, and hence gives a higher 
value of $\Omega_{\rm gas}$.  For a range of $J_{912}$ and $C$, we find that 
$\approx$ 5--20\% of the total baryons predicted by BBN are in extended gaseous
envelopes of galaxies.

\end{document}